# Big Data Model "Entity and Features"

*N. Shakhovska, U. Bolubash, O. Veres*

*Lviv Polytechnic National University ;*
*e-mail: Natalya.B.Shakhovska@lpnu.ua,*
*Oleh.M.Veres@lpnu.ua*
*bol_jura@ukr.net*

*Abstract*. The article deals with the problem which led to Big Data. Big Data information technology is the set of methods and means of processing different types of structured and unstructured dynamic large amounts of data for their analysis and use of decision support. Features of NoSQL databases and categories are described. The developed Big Data Model "Entity and Features" allows determining the distance between the sources of data on the availability of information about a particular entity. The information structure of Big Data has been devised. It became a basis for further research and for concentrating on a problem of development of diverse data without their preliminary integration.

*Key words*: Big Data, NoSQL, document-oriented database, Big Table.

## INTRODUCTION

The term Big Data was introduced by Clifford Lynch, editor of *Nature*, who released a special issue on 3 September 2008 examining what big data sets meant for modern science. He collected information about the phenomenon of explosive growth and diversity of data and technological prospects in the paradigm of probable transition from quantity to quality [1].

Big data is the term increasingly used to describe the process of applying serious computing power – the latest in machine learning and artificial intelligence – to seriously massive and often highly complex sets of information (cited from 4/2013 the Microsoft Enterprise Insight Blog).

Typically Big Data:
- is automatically machine obtained/generated,
- may be a traditional form of data now expanded by frequent and expanded collection,
- may be an entire new source of data,
- is not formatted for easy usage,
- can be mostly useless, although Big Data is collected and its economics is positive,
- is more useful when connected to structured data in corporate enterprise systems (ERPs).

The challenges include capture, curation, storage, search, sharing, transfer, analysis, and visualization

Big Data has many advantages over traditional structured databases. The properties of Big Data enable analysis for the purpose of assembling a picture of an event, person, or other object of interest from pieces of information that were previously scattered across disparate databases. Big Data is a repository for multi-structure data which makes it possible to draw inferences from correlations not possible with smaller datasets.

Despite the fact that the term was introduced in the academic environment, the primary problem was the growth and diversity of scientific data in practical tasks. As of 2009, the term was widely used in the business press, and 2010 saw emergence of the first series of products and solutions related only to the problems of processing of huge data volumes. By 2011, most of the largest providers of information technology for organizations based their business strategies on the concept of big data, including IBM, Oracle, Microsoft, Hewlett-Packard and EMC [1].

Problems arising during processing, interpretation, collection and organization of Big Data appeared in numerous sectors, including business, industry and non-profit organizations. Data sets such as retail customer transactions, weather monitoring, business analysis can quickly outstrip the capacity of the traditional methods and tools for data analysis. There are new methods and tools such as databases NoSQL and map Reduce, natural language processing, machine learning, visualization, acquisition, and serialization. It is necessary to fully understand what happens when we use growing big data and where its role is becoming crucial. Knowing the requirements to the existing methods of system development and data analysis is also important.

## ANALYSIS OF RECENT RESEARCHES AND PUBLICATIONS

One of the adapting concepts not only of relational data is NoSQL. Followers of the concept of NoSQL language emphasize that it is not a complete negation of SQL and the relational model, and that the project comes from the fact that SQL is an important and very useful tool, which, however, cannot be considered universal. One of the problems mentioned for classical relational database is the problem of dealing with huge data, projects with a high load and parallel processing. The main objective of the approach is to extend the database if SQL is not flexible enough, and not displace it wherever it performs its tasks.

The NoSQL idea is underlain by the following points:
• non-relational data model,
• distribution,

• open output code,
• good horizontal scalability.

As one of the methodological approaches of NoSQL studies, a heuristic principle is used, known also as the CAP theorem (Consistence, Availability, Partition tolerance – «consistency, availability, resistance to division"), arguing that in a distributed system consistency, accessibility (English: availability, every query is responded to) and resistance to splitting a distributed system into isolated parts cannot be simultaneously provided. Thus, if it is necessary to achieve high availability and stability of the division, the means to ensure consistency of data provided by traditional SQL-oriented database with transactional mechanisms on the principles of ACID are not to be focused on [1].

A non-strict proof of the CAP theorem is based on a simple reflection. Let the distributed system consist of $N$ servers, each of which handles the requests of a number of client applications. While processing a request, the server must ensure the topicality of the information contained in the response to the request being sent, which previously required synchronizing the contents of its own base with other servers. Thus, the server must wait for a full synchronization or generate a response based on non-synchronized data. In the alternative case, for some reasons synchronization involves only some of the servers. In the first case the requirement of availability is not fulfilled, the second one fails to satisfy the requirement of consistency and in the third case resistance to division requirement is not matched.

There are four categories of NoSQL databases.

The first category is *Key-value stores*. These are very simple stores. Actually, they are large cache tables, where each key has its value. Such databases are able to quickly process a huge volume of information, but they are limited in terms of query language (just searching for a key or value). The examples of key-value databases are Dynomite, Voldemort, Tokyo, Redis, etc.

The second category is *Bigtable clones*. Bigtable is a database developed by Google for its own needs. The database is a large three dimensional table comprising columns, lines, and time markers. Such architecture allows it to obtain very high productivity and, moreover, can be easily scaled on many computers. However, it is a non-relational database which does not support many features peculiar to relational databases. In particular, Bigtable lacks complex queries, joining operations function, etc. Google does not promote Bigtable. That is why there are several independently developed clones of this database on the market, including such projects as Hadoop, Hypertable, and Cassandra.

The next category of bases is the *document-oriented databases*. These bases are partially similar to Key-Value bases, but in this case a database knows what constitutes value. Typically, the value is a document or object to the structure of which queries can be made. Examples of such bases are CouchDB and MongoDB.

The fourth category is *databases based on graphs*. These bases are targeted at supporting complex relationships between objects and are based on graph theory. The data structure in such databases is a set of nodes connected by links. In this case nodes and links can have a number of attributes. Neo4j, AllegroGraph and Sones graphDB are the examples of such databases.

There is also the fifth category, but it is not considered to be of NoSQL type. These are *object-oriented databases*. Such databases serve first of all to maintain object-oriented programming paradigms. It is extremely easy to use them in programming languages supporting this paradigm. There are several mechanisms to access data in a NoSQL database.

**Restful interfaces**. This is an interface similar to HTTP, the main Internet protocol. Within the framework of this approach, each object that can be manipulated is supposed to have its own unique address. Using this address, the marked object can be queried, created, edited and deleted. Meanwhile, no state is stored on the server, which means each query is processed independently of other queries.

**The query language different from SQL**:
- GQL – Sql-similar language for Google Bigtable,
- SPARQL – semantic Web query language,
- Gremlin – graph traversal language,
- Sones Graph Query Language – Sones Graph query language.

**API requests**:
- Google Bigtable Datastore API,
- Neo4j Traversal API.

NoSQL can offer a high level of operational readiness, correctness and productivity. Apparently, the main advantage of NoSQL database is productivity. All NoSQL databases surpass relational databases in its niche. If until recently there was only one type of database in all cases – relational database, today the situation has changed. For each particular case, one has to select its data warehouse. Sometimes one has to have several databases simultaneously, in each of which its strongest points are used. For example, in web-applications Mongodb is used as the main data warehouse, and with the help of Redis the user's query caching is organized. As a result, we obtain a very high performance system with a developer-friendly interface. Another important advantage of NoSQL databases is that many representatives of this family of data warehouses are implemented as projects with open source. In general, comparison of SQL and NoSQL is presented in Fig. 1.

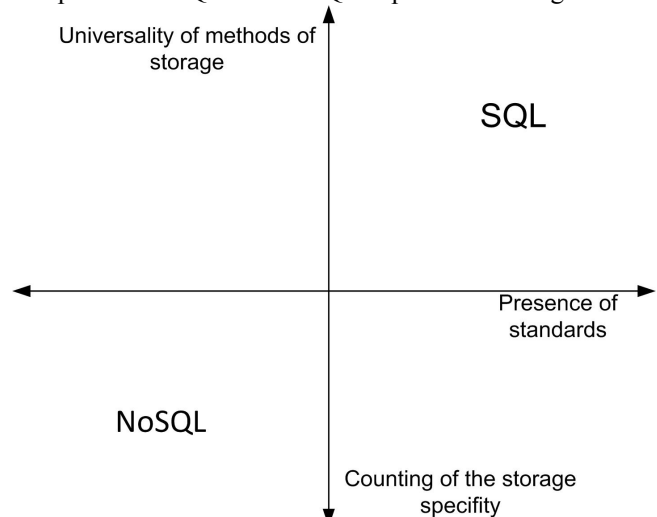

**Fig.1.** Comparison of SQL and NoSQL

Thus, the existence of different categories of NoSQL databases requires the formal description of data models which are processed by them.

Another comparison of relational database and NoSQL-system is presented in Table 1.

**Table 1.** Comparison of relational DB and NoSQL-system

| Relational DB | NoSQL-system |
|---|---|
| Structured data | Unstructured data |
| ACID | Without ACID |
| Strong data consistence | Fractional data consistence |
| ETL | Without ETL |
| Not so fast response | Fast response |
| Efficiency | Flexibility |

The Big Data issues are still not very well delineated, although it is the centre of gravity for business and technology. Analysis of the above-mentioned sources, popular science magazines, and blogs allow identifying the following discussion focuses:
- sources of big data;
- hardware and infrastructure;
- software and storage;
- IT (methods and tools of data processing);
- using big data, business-analysis.

Devices and people can be singled out as sources of data. Examples of the former include national and international projects such as Lange Hadron Collider in CERN, the European Laboratory for Particle Physics, the Large Synoptic Survey Telescope in the north of Chile, Internet things, industry (SCADA, finance, etc.).

The second type of data sources are represented by social networks, health care, retail, personal location data, public sector management, etc.

For data collection and processing it is practicable to use cloud computing technology. Cloud computing is a new paradigm for placing clusters of data and providing different services through a local network or via Internet. Hosting of date clusters allows clients to keep and calculate huge amount of data in a cloud.

On the one hand, when we collect big data, we have an opportunity to support decisions with the help of BI. BI is a set of theories and technologies aimed at data transfer of relevant and useful information for business processes. For example, according to the analyst Tim Swanson, the number of operations made in Cryptocurrency Bitcoin is more than 100 000 [12]. According to the IDC Digital Universe Study, the total amount of global data in 2005 was 130 Exabyte, by 2011 it rose to 1227 EB, and during the 2014 tripled and reached 4.4 ZB (zettabyte is $10^{21}$ bytes). A forecast made by the same study shows that by 2020 the volume of digital data will increase up to 44 ZB (the annual increase by 40 %) [13].

The size of an individual database increases as fast as that and has passed the petabyte barrier, for instance, for social networks databases. Therefore, online processing of such volumes in a distributed mode is practically impossible (Fig. 2).

Table 2 [1] shows some tools with processing of Big Data with open output code, which are provided by the cloud computing infrastructure. Most of the tools are provided by Apache and produced under the Apache license. These products are grouped depending on the tasks that arise during the processing of big data.

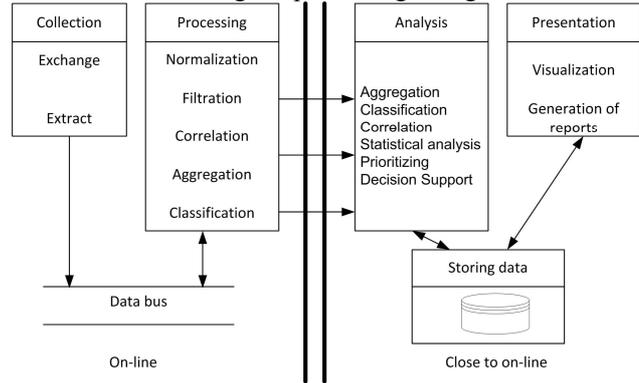

**Fig.2.** Contrastive description of OLAP and Big Data

**Table 2.** Tools for working with Big Data [1]

| Tools for Big Data | Description |
|---|---|
| *Data analysis tools* | |
| Ambari http://ambari.apache.org | Web tool for service delivery, management and monitoring of Apache Hadoop clusters. |
| Avro http://avro.apache.org | The system of data serialization. |
| Chukwa http://incubator.apache.org/chukwa | The system for collecting data to manage large distributed systems. |
| Hive http://hive.apache.org/ | Data warehouse infrastructure that provides data aggregation. |
| Pig http://pig.apache.org | High-level data streams language and executable framework for parallel computing. |
| Spark http://spark.incubator.apache.org | Fast and general Hadoop data computer/calculator. It provides a simple and expressive programming model that supports a wide range of applications, including ETL, machine learning, flows processing. |
| ZooKeeper http://zookeeper.apache.org/ | Highly productive coordinating service for distributed applications |
| Actian http://www.actian.com/about-us/#overview | Provides storage of raw data and prepares the data for further analysis |
| HPCC http://hpccsystems.com | Provides rapid transformation, parallel processing for use with Big Data |
| *Tools of Data Mining* | |
| Orange http://orange.biolab.si | Open source data visualization and analysis for novice and experts. |
| Mahout http://mahout.apache.org | Library of facilities of machine learning and data mining |
| KEEL http://keel.es | Evolutionary algorithm for data mining problems |
| *Social Networking Tools* | |
| Apache Kafka | Platform high bandwidth for data processing in real time |
| *Tools of BI* | |
| Talend http://www.talend.com | Data integration, management, integration of applications, tools and services for Big Data |
| Jedox http://www.jedox.com/en | Functions of analysis, reporting, planning |
| Pentaho http:// | Data integration, business |
| www.pentaho.com | analysis, data visualization, prediction |
| Rasdaman http://rasdaman.eecs.jacobs-university.de/ | Multidimensional raster data (array) without restrictions on size, availability of query language |
| *Search tools* | |
| Apache Lucene http://lucene.apache.org | Applications for full-text indexing and search |
| Apache Solr http://lucene.apache.org/solr | Full-text search, faceted search, dynamic clustering, formats of document of type Word, PDF, spatial search |
| Elasticsearch http://www.elasticsearch.org | Distributed text search tool with a web interface and JSON documents |
| MarkLogic http://developer.marklogic.com | NOSQL and XML database |
| mongoDB http://www.mongodb.org | Cross-platform document-oriented database management system with support for JSON and dynamic schemes |
| Cassandra http://cassandra.apache.org | Scalability and high availability without compromising performance. |
| HBase http://hbase.apache.org | Is the Hadoop database, a distributed, scalable, big data store. |
| InfiniteGraph http://www.objectivity.com | Distributed Graph Database |

## OBJECTIVES

Big data is a term used to identify data sets that we cannot cope with using existing methodologies and software tools because of their large size and complexity. Many researchers are trying to develop methods and software tools for data mining or information granules of Big Data.

Big Data features are: work with unstructured and structured information; targeting at faster data processing; leading to the fact that traditional query languages are ineffective while working with data.

The purpose of the article is to formally describe different data models, operations and carriers distinguishing and sharing methods since traditional query languages are ineffective for working with data.

## A FORMAL DESCRIPTION OF THE BIG DATA STRUCTURE

A striking example of Big Data is a data set that describes functioning of a region. Therefore, in the end there are:

- a large set of entities: persons, places, organizations (individual and legal), date, natural resources (rivers, forests, lakes), recreational resources (historical monuments, health care), legislative acts and reports,
- huge database features: documents for data mining, ontological terms, data dictionaries, which allow associating certain objects.

On the basis of this information the relations between entities should be established.

Formally, all the objects fall under the following categories [1]:

- $e$ – entities,
- $f$ – features,
- associations between entities $e$ and features $f$.

For instance:
- name $e$ is mentioned in $f$ document,
- notion $f$ appeared in $e$ document.

There are also defined:
- set $E$ of entities,
- set $F$ of features,
- for each $e$ and $f$ the number of associations between $e$ and $f$ is designated as $n_{e,f}$.

The total number of entities is determined as $|E|$, the total number of features is determined as potency of sets $F$: $|F|$. Let us also describe:

- for every feature $f$ plural $e(f) = \{e \in E : n_{e,f} > 0\}$ of all entities associated with $f$,
- for every entity $e$ plural $f(ef) = \{f \in EF n_{e,f} > 0\}$ of all features associated with $e$.

Let us describe these qualitative representations in the quantitative form.

In similar situations when a few entities are related to a feature, we will use the quantitative representation of information, i.e. the number of binary questions (yes, no) required to find the object we need. In general, if we know that the unknown object belongs to a set consisting of $N$-elements, this set can be divided into halves.

Therefore, the number of objects will be $\frac{N}{2}$. Let us continue this procedure: the second question is asked, for which we will divide the selected half into halves. Thus, after two questions (actions) we will have $\frac{N}{4}$ objects to which the unknown one belongs. After three questions (actions) we will receive $\frac{N}{8}$. Answering $q$ binary questions will result in a set of $N \cdot 2^{-q}$ elements that contain the necessary object. Thus, for $N$ alternatives the relevant information (the number of binary questions) is $N \cdot 2^{-q} = 1$ and therefore equal to $q = \log_2(N)$.

Entities can be described in a similar way. There is $|E|$ entity with $\log_2(|E|)$ amount of information. When we know that some entity is associated with a response (we have $|e(f)|$ entity), then the amount of information is equal $\log_2(|e(f)|)$. Therefore, the fact of the relation between the entity and the feature $f$ allows reducing the number of questions to $\log_2(|E|) - \log_2(|e(f)|) = \log_2\left(\frac{|E|}{|e(f)|}\right)$. It is similar to the formula defining amount of information.

Besides, the effect of several associations can be described by counting how many additional binary questions we can ask to further know an association with the required entity. Let us start with $n_{e,f}$. Each binary question reduces the number of objects by half; $q$ questions reduce the number to $n_{e,f} \cdot 2^{-q}$. We continue to have an association up to the point at which the number of objects becomes $\geq 1$. Most $q$ for which we still do not have an association is defined as $n_{e,f} \cdot 2^{-q} = 1$, which, in its turn, is defined as $q = \log_2(n_{e,f})$. Adding an additional question is defined as $1 + \log_2(n_{e,f})$.

General importance of $f$ features for the entity $e$ is defined as $\log_2\left(\frac{|E|}{|e(f)|}\right)$ with the factor of importance $1 + \log_2(n_{e,f})$. The resulting amount of information is defined as:

$$I(e,f) = \left(1 + \log_2(n_{e,f})\right) \cdot \log_2\left(\frac{|E|}{|e(f)|}\right). \quad (1)$$

This formula (1) is one of the options in terms of frequency – the so-called inverse document frequency *tf-idf*. For each *e*-entity we have importance $I(e, f)$ for different features *f*. It is necessary to normalize the meaning of importance, which looks like cosine normalization:

$$V(e,f) = \frac{\left(1 + \log_2(n_{e,f})\right) \cdot \log_2\left(\frac{|E|}{|e(f)|}\right)}{\sqrt{\sum \left(\left(1 + \log_2(n_{e,f})\right) \cdot \log_2\left(\frac{|E|}{|e(f)|}\right)\right)^2}}. \quad (2)$$

For each *e*-entity there is weight $V(e,f)$. Thus, as a measure of proximity between two objects $E_1$ and $E_2$, we can consider the distance between the corresponding vectors $(V(e,f_1), V(e,f_2), ...)$.

In the ordinary Euclidean distance $d(a,b) = \sqrt{(a_1 - b_1)^2 + ...}$ squares of differences are added. Thus, for each weight $V(e,f)$, which represents the number of bits, we will have the following equation:

$$d(e_1, e_2) = \sum_{f \in F} |V(e_1, f) - V(e, f_2)|. \quad (3)$$

This distance depends on the number of features: for example, if in addition to the documents we store their copies, the distance is doubled. In order to avoid this dependence, the distance $d(e_1, e_2)$ is ordinarily normalized in the range [0,1] through the division by the maximum possible value of this distance.

How can we estimate the largest possible value of this distance? In general, when we do not know the true value of A and B**,** the two non-negative quantities, and we know only the upper limits of these variables $\bar{a}$, $\bar{b}$, then the largest possible value of the difference $|\bar{a} - \bar{b}|$ is equal to $\max(\bar{a}, \bar{b})$. Then [2, 16]:

- if $\bar{a} \leq \bar{b}$, then $|\bar{a} - \bar{b}| = \bar{b} - \bar{a} \leq \bar{b}$ that is why, $|\bar{a} - \bar{b}| \leq \max(\bar{a}, \bar{b})$;
- if $\bar{b} \leq \bar{a}$, then $|\bar{a} - \bar{b}| = \bar{a} - \bar{b} \leq \bar{a}$ that is why, $|\bar{a} - \bar{b}| \leq \max(\bar{a}, \bar{b})$.

In both cases we have $|\bar{a} - \bar{b}| \leq \max(\bar{a}, \bar{b})$.

The limit $\max(\bar{a}, \bar{b})$ is reached:
- when $\bar{a} \leq \bar{b}$, if $a = 0, b = \bar{b}$;
- when $\bar{b} \leq \bar{a}$, if $a = \bar{a}, b = 0$.

## IV. ASSOCIATION MODELS OF ENTITIES AND FEATURES FOR VARIOUS CATEGORIES OF NOSQL DATABASE

We introduce the concept of model associations between objects and features for different categories NoSQL databases.

The data carrier in the model «key-value», also known as column DB, is described with cortege in the following way:

$$KV = \{<f, e>\}, \quad (4)$$

where: $f$ is the key that takes a unique value in each pair; $e$ is the value that corresponds to this key. Keys can be folded (major or minor), and the value supports unlimited semantics.

The model signature has the following form:

$$O = \langle \pi, \sigma \rangle, \quad (5)$$

where: $\pi$ is an operation projection by attributes (key or value); $\sigma$ is a selection of attributes (value selection by key, keys by value, key by ancestors value). These operations refer to the reading category [5, 6].

An example of the column database is **Cassandra**.

The model is used in the system BigTable of Google and was designed for the distributed storage of large volumes of data:
- not a full relational data model,
- dynamic control support of data placement.

The BigTable data model is simple and contains rows, columns and timestamps:

$$BigTable = \{<r, c, t>\}. \quad (6)$$

The address of the documents from the Internet may be presented as the row names in the database of the search engine, and the features of these documents can serve as columns' names (for instance, the content of the document can be stored in column «Content» and the reference to secondary pages in «Anchor»).

Another example is Google Maps that consist of billions of images, and each of the image details a certain geographic area of the Earth. The structure of Google Maps in BigTable is based on the fact that each row corresponds to a single geographic segment, and the columns are the images, from which this segment is built; different columns have the images with different resolutions.

If one-type data is stored in several columns, such columns make a family, due to the BigTable model, which takes the following form:

$$colF = \{c_i, c_j \mid dom(c_i) \in T \wedge dom(c_j) \in T\} \quad (7).$$

A column family can be first of all used for compressing similar data to decrease volume. A column family is a unit of data access.

The BigTable rows are also important (they can be 64 kilobytes in length). An operation of tag to row is atomic, meaning that no other programme can change the data in the column family of the row until the tag to row of the previous programme is succeeded).

In addition, the rows are easily sortable. An example of the URL-document, after its record has been made reversible, shows how easily all lines are organized as a third level domain name.

The content of web-pages is constantly changing. In order to accommodate these changes, each copy of the data stored in the column is given a time stamp. In BigTable, timestamp is a 64-bit number that can code the date and time as required. For example, a timestamp for copies of web-pages in the column *Contents* will be the creation date and time of such copies. Using timestamps, applications can specify a search in BigTable, for example, of only recent data copies.

Therefore, for the domain of any Google service, its own data map BigTable can be created, which contains any number of rows and a set column families unique for this domain. The inevitable repetitions of data in columns are sorted by timestamps. All this appoints to the complete lack of support of ACID properties.

However, the main advantage of this approach is that such database can be easily divided into independent parts and distributed to a set of servers. Alphabetically sorted lines are shared on ranges called Tablets, or dependent tables. As lines are sorted by a key name in every tablet, it is very simple for client applications to find either the necessary tablet or the necessary line in it.

In this model the key identifies the line containing the data which is stored in one or several sets of columns. Every line can have many values of columns within such sets. The value of each column contains a time tag, which is why some values of compliances between line and column can be within one family of columns.

BigTable is a big and distributed system for such synchronizing objects, due to which a distributed lock service that Google call Chubby has been used instead. Its role in BigTable can be compared with a role of transactions in the usual Database Management System (DBMS). For each tablet-server, Chubby creates a special chubby-file. Due to this file, BigTable Master is always aware of efficient servers. One more chubby-file contains links to the location of the Root tablet with data on the location of all the other Tablets. This file informs the Master which servers are managed by the Tablet.

Undoubtedly, use of Chubby-service in BigTable to some extent solves the problem of data consistency control with a set of remarks in the distributed environment. However, consistency can be of different kinds. BigTable became the first attempt to reach the balance between system productivity, its scalability and consistency of data stored in it. This resulted in the maintenance of the so-called weak consistency, which, in principle, met the requirements of major services working with BigTable.

Fig. 3 shows how the user looks for the tablet [4].

The carrier of object-document model is described by the following cortege:

$$OD = \left\langle f_0, \langle f_1 : e_1, f_2 : e_2, ..., f_n : e_n \rangle, \langle f_1 : d_1, f_1 : d_2, ..., f_{n+l} : d_l \rangle \right\rangle, \quad (8)$$

where: $f_0$ is the identifier of the document; $f_1..f_n$ are the attributes of the document; $e_1..e_n$ are the atomic values of features $f_1..f_n$; $d_1..d_l$ are dispatchers to the other documents, $d_i = e(f_i)$.

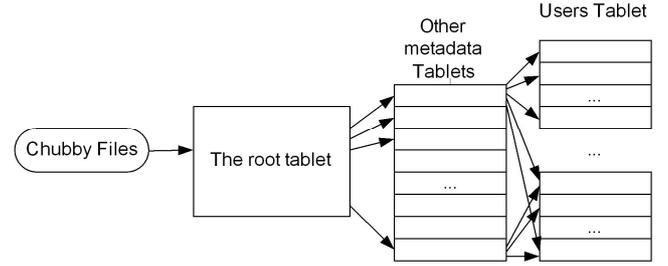

**Fig. 3.** The hierarchy of tablets

Operations of this model are object.

The operation of determination of element nodes will look as follows:

$$v(f_i) = \{C\} \cup \{od_i \mid i = \overline{1,n}\} \cup \{e(f_i) \mid i = \overline{0,n+l}\}, \quad (9)$$

where: $C$ is a collection of documents $od_i$

The operation of determination of node values will have the following form:

$$v(f_i) = \{e_{ij} \mid i = \overline{1,n}, j = \overline{0,m+l}\}, \quad (10)$$

where: $e_{ij}$ is the value of attribute $f_i$.

The relations between carrier elements are also defined.

The relation *element-element* is defined between documents and collection:

$$OD \times C \to EE . \quad (11)$$

Relations *element-attribute* look as follows:

$$f_i \times OD \to EA . \quad (12)$$

Relations *element-tag* are expressed in the following way:

$$f_i \times d_j \to ER . \quad (13)$$

The relations *element-data* are defined as follows:

$$f_i \times e_j \to ED . \quad (14)$$

MongoDB and CouchDB are examples of this type of DBMS.

The graph data model is presented as:

$$O = \langle ID, A, z, r \rangle, \quad (15)$$

where: *ID* is a set of identifiers, graph nodes; *A* is a set of labelled directed arcs $(p, l, c), p, c \in ID$; *l* is line-tag; the record $(p, l, c)$ means that there is a relation *l* between nodes *p* and *c*; *z* is a function that displays each node $n \in ID$ in the specific composite or atomic value, $z : n \to v$; *V* is a special root graph node.

The structure of the XML document which consists of the enclosed element-tags is well-known. Its difference from the graph model considered above generally consists in interpretation of tags: in column tags are used as designation of communications between elements of data schemes and tags are not necessary for designation of an element, and in XML document-focused model it is necessary that each (not text) element of data has an identifying sign. XML is also translated into the tree data structure, which is a specific case of graph model.

In the XML graph model, semistructured data requires specialized types of attributes, such as *ID*, *IDREF* and *IDREFS*. The specified types enable organizing the storage of cross tags in *XML*-elements such as <*eid*, *vahie*=""> (<*the element identifier*, *value*>) and attributes of <*label*, *eid*=""> (<*tag*, *value*>) type.

There are several types of *RDF* data as a graph model, including *RDF/XML, N3, Turtle, RDF/JSON.*

The description of resources in the form of *RDF*-Data sets are defined as a triplet "subject" - "predicate" - "object", that is for the set *U* (*Universal Resource Identifier, URI*, unified identifier of resources) these are elements *f*, for the set *B* (*Black nodes, empty nodes*), set *L* (*Literal*, *RDF*-literals), $B \in e, L \in e$ the set is defined as (*f, e(f), e*), where *f* is "subject", *e(f)* is "predicate" and *e* is "object".

For *RDF*-graph model of data, let $t = (f, e(f), e)$ be an *RDF*-element of data, where $(f,e(f),e) \in (UB) \times U \times (UBL)$; besides, *t* is a key element if it does not contain nodes without identifiers. *RDF*-graph $G$ is a $T \supseteq t$ set [3].

Consequently, Big Data includes various models of data [18 – 22]. Therefore, there should be methods of their transformation with the minimum loss of data.

Information structure of Big Data is presented in Fig.4.

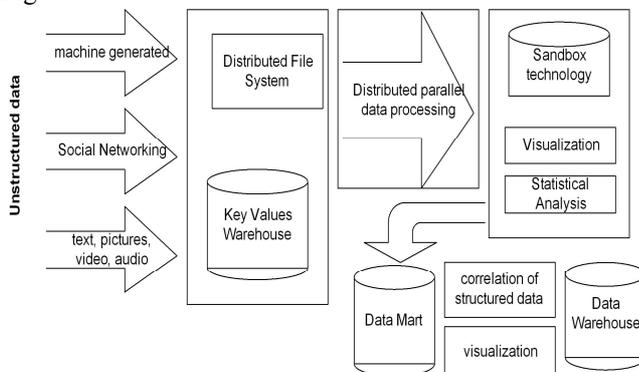

**Fig. 4**. Information structure of Big Data.

CONCLUSIONS

The article discusses the structure of Big Data. Models of object associations and features of the main data presentations in NoSQL are defined. The information structure of Big Data has been devised. It became a basis for further research and for concentrating on a problem of development of diverse data without their preliminary integration.

REFERENCES

1. **Pedrycz W., and S.-M. Chen. 2015.** Information Granularity, Big Data, and Computational Intelligence, Studies in Big Data 8, DOI: 10.1007/978-3-319-08254-7, Springer International Publishing Switzerland.
2. **Srinivasa, S., and V. Bhatnagar. 2012.** Big data analytics. In: Proceedings of the First International Conference on Big Data Analytics BDA'2012. Lecture Notes in Computer Science, vol. 7678. Springer, New Delhi, 24–26 Dec 2012.
3. **Butakova M.A., Klimanskaya E.V., and Yants V.I. 2013.** Measure of informative similarity for the analysis of semi-structured information. Modern problems of science and education. n. 6; Available online at: <http://www.science-education.ru/113-11307>.
4. **Chang F., Dean J., Ghemawat, S., Hsieh W., Wallach D., Burrows M., Chandra T., Fikes A., Gruber R. 2006.** Bigtable: A Distributed Storage System for Structured Data, Research (PDF), Google.
5. **Papakonstantinov F. and Widom J. 2005.** Object exchange across heterogeneous information sources. 11-th International Conference on Data Engineering (ICDE'05). 251-261.
6. **Feng Z., Hsu W. and Li M. 2005.** Efficient pattern discovery for semi-structured data. 17th IEEE International Conference on Tools with Artificial Intelligence (ICTAI-05). 301-309.
7. **Shakhovska N. 2010.** Software and algorithmic software of data storage and data spaces: monograph; Acting Lviv Polytechnic Nat. University. Lviv: Izd Lviv Polytechnic, 194. (in Ukrainian).
8. **Shakhovska N. 2008.** Features Data spaces modeling. Proceedings of Lviv Polytechnic National University. Lviv: Izd Lviv Polytechnic Nat. Univ, Computer Engineering and Information Technology. 608:145 – 154. (in Ukrainian).
9. **Shakhovska N., Medykovsky M., and Stakhiv P. 2013.** Application of algorithms of classification for uncertainty reduction/ Przeglad Elektrotechniczny. 89(4):284-286
10. **Кut V. 2014.** Consolidated model of remote and consulting center of people with special needs. Proceedings of Lviv Polytechnic National University. Information systems and networks. 783:120–127. (in Ukrainian).
11. **Kushniretska I., Kushniretska O., and Berko A. 2014.** Analysis of information resources integration system of dynamic semistructured data in a Web-environment. Proceedings of Lviv Polytechnic National University. Information systems and networks. 805:162-169. (in Ukrainian).
12. The daily volume of transactions in Bitcoin overcame the barrier of 100 000: Available online at: <http://vkurse.ua/ua/business/ezhednevnyy-obem-tranzakciy-v-bitcoin.html>.
13. Data Growth, Business Opportunities, and the IT Imperatives Available online at: <http://www.emc.com/leadership/digital-niverse/2014iview/executive-summary.htm>.
14. **Shakhovska N., Bolubash Yu, and Veres O. 2015.** Big Data Federated Repository Model. Proc. of CADMS'2015. Lviv: Lviv Polytechnic Publishing. 382-384. (in Ukrainian).
15. **Shakhovska N., Bolubash Y, and Veres O. 2014.** Big data organizing in a distributed environment. Computer Science and Automation. Donetsk. Ukraine. 2(27):147-155. (in Ukrainian).
16. **Shakhovska N. 2011.** Formal presentation of data space as an algebraic system. System Research and Information Technologies. National Academy of Sciences of Ukraine, Institute for Applied Systems Analysis. Kyiv. 2:128 – 140. (in Ukrainian).
17. **Shakhovska N. and Bolubash Y. 2013.** Working with Big Data as indicators of socio-ecological-economic development. Eastern-European Journal of Enterprise Technologies, 2(65):5. (in Ukrainian).
18. **Kalyuzhna N. and Golovkova К. 2013.** Structural contradictions in control system by enterprise as

function of associate administrative decisions. EconTechMod: an International Quarterly Journal on Economics in Technology, new Technologies and Modelling Processes. Krakiv-Lviv, 2(3):33-40.
19. **Magoulas R., and Ben L. 2009** Big data: Technologies and techniques for large scale data, Release 2.0
20. **Kossmann D., Dittrich J.P. 2006** Personal Data Spaces. Available online at: http://www.inf.ethz.ch/news/focus/res_focus/feb_2006/index_DE.
21. **Stonebraker M., Abadi D., DeWitt D. J., Madden S., Pavlo A., and Rasin A. 2012** MapReduce and Parallel DBMSs: Friends or Foes. Communications of the ACM (53:1). 64-71.
22. **Laney D. 2001** 3D Data Management: Controlling Data Volume, Velocity and Variety. Gartner.